\documentclass[10pt]{llncs}
\usepackage{graphics,amssymb,amsmath,epsfig}
\usepackage{longtable}
\makeatletter

\def\vp{{\rm VP}}
\def\vnp{{\rm VNP}}

\def\per{{\rm per}}
\def\ham{{\rm ham}}
\def\p{{\rm P}}
\def\np{{\rm NP}}
\def\sharpp{{\rm \sharp P}}

\makeatletter

\begin{document}

\title{On the expressive power of
permanents and perfect matchings
of matrices of bounded pathwidth/cliquewidth
}

\author{Uffe Flarup
\inst{1},
Laurent Lyaudet
\inst{2}
}
\institute{Department of Mathematics and Computer Science\\
Syddansk Universitet,
Campusvej 55, 5230 Odense M, Denmark\\
e--mail: flarup@imada.sdu.dk; fax: +45 65 93 26 91\\
\and Laboratoire de l'Informatique du Parall\'elisme\footnote{UMR 5668 ENS Lyon,
  CNRS, UCBL, INRIA. Research Report RR2008-05}\\
Ecole Normale Sup\'erieure de Lyon,
46, all\'ee d'Italie, 69364 Lyon Cedex 07, France\\
e--mail: laurent.lyaudet@ens-lyon.fr; fax: +33 4 72 72 80 80
}

\maketitle

\begin{abstract}
Some 25 years ago Valiant introduced an algebraic model of computation
in order to study the complexity of evaluating families of
polynomials. The theory was introduced along with the complexity classes
$\vp$ and $\vnp$ which are analogues of the classical classes $\p$
and $\np$.
Families of polynomials that are difficult to evaluate (that is,
$\vnp$-complete) includes the permanent and hamiltonian polynomials.

In a previous paper the authors together with P. Koiran studied the expressive power of
permanent and hamiltonian polynomials of matrices of
bounded treewidth, as well as the expressive power of perfect
matchings of planar graphs.
It was established that the permanent and hamiltonian polynomials of
matrices of bounded treewidth are equivalent to arithmetic formulas.
Also, the sum of weights of perfect matchings of planar graphs was shown
to be equivalent to (weakly) skew circuits.

In this paper we continue the research in the direction
described above, and study the expressive power
of permanents, hamiltonians and perfect matchings of matrices
that have bounded pathwidth or bounded cliquewidth.
In particular, we prove that permanents, hamiltonians and perfect matchings of matrices
that have bounded pathwidth express exactly arithmetic formulas. 
This is an
improvement of our previous result for matrices of bounded treewidth.
Also, for matrices of bounded weighted cliquewidth we show membership
in $\vp$ for these polynomials.
\end{abstract}

\section{Introduction}
In this paper we continue the work that was started in \cite{FKL}.
Our focus is on easy special cases of otherwise difficult to evaluate
polynomials, and their relation to various classes of arithmetic circuits.
It is conjectured that the permanent and hamiltonian polynomials are hard
to evaluate. Indeed, in Valiant's model \cite{Val79,Val82} these families of polynomials
are both $\vnp$-complete.
In the boolean framework they are complete for the complexity class
$\sharpp$ \cite{Val79a}. However, for matrices of bounded treewidth the
permanent and hamiltonian polynomials can efficiently be evaluated - the
number of arithmetic operations being polynomial in the size of the
matrix \cite{CMR}.

An earlier result along these lines is related to computing weights
of perfect matchings in a graph:
The sum of weights of all perfect matchings in a weighted (undirected) graph
is another hard to evaluate polynomial, but for planar graphs it can be
evaluated efficiently due to Kasteleyn's theorem \cite{Ka}.

By means of reductions these evaluation methods
can all be seen as general-purpose
evaluation algorithms for certain classes of polynomials.
As an example, if an arithmetic formula represents a polynomial $P$ then
one can construct a matrix $A$ of bounded treewidth such that:

\begin{itemize}
\item[(i)] The entries of $A$ are variables of $P$, or constants from
the underlying field.
\item[(ii)] The permanent of $A$ is equal to $P$.
\end{itemize}

It turns out that the converse holds as well, so with respect
to the computational complexity computing the permanent
of a bounded treewidth matrix is equivalent to evaluating an arithmetic
formula.
In \cite{FKL} the following results (with abuse of notation)
were established:

\begin{itemize}
\item[(i)] permanent/hamiltonian(bounded treewidth matrix) $\equiv$ arithmetic formulas.
\item[(ii)] perfect matchings(planar matrix) $\equiv$ arithmetic skew circuits.
\end{itemize}

One can also by similar techniques show that:
\begin{itemize}
\item[(iii)] perfect matchings(bounded treewidth matrix) $\equiv$ arithmetic formulas.
\end{itemize}

Other notions of graph ``width'' have been defined in the litterature
besides treewidth, e.g. pathwidth, cliquewidth and rankwidth.
Here we would like to study the evaluation methods mentioned above,
but considering matrices $A$ that
have bounded pathwidth or bounded cliquewidth instead of bounded treewidth.
In this paper we establish the following results:

\begin{itemize}
\item[(i)] per/ham/perf. match.(bounded pathwidth matrix) $\equiv$ arithmetic skew
circuits of bounded width $\equiv$ arithmetic weakly skew circuits of bounded width $\equiv$
arithmetic formulas.
\item[(ii)] arithmetic formulas $\subseteq$ per/ham/perfect matchings(bounded
cliquewidth matrix)
$\subseteq$ $\vp$.
\end{itemize}


\emph{Overview of the paper.} The second section of the paper introduces definitions used throughout the paper
and provides some small technical results related to graph widths.
In particular we
show equivalence between the weighted definitions of cliquewidth, NLC-width and
m-cliquewidth with respect to boundedness.
Sections 3 and 4 are devoted to the
expressiveness of the  permanent, hamiltonian, and perfect matchings of the
graphs of bounded pathwidth and bounded weighted cliquewidth respectively. 
We prove in Section 3 that permanent, hamiltonian, and perfect matchings limited
to bounded pathwidth graphs express arithmetic formulas. In Section 4, we show
that for all three polynomials the complexity is between arithmetic formulas and
$\vp$ for graphs of bounded weighted cliquewidth.

\section{Definitions and preliminary results}

\subsection{Arithmetic circuits}

\begin{definition}
An {\em arithmetic circuit} is a finite, acyclic, directed graph. Vertices have
indegree 0 or 2, where those with indegree 0 are referred to as {\em inputs}.
A single vertex must have outdegree 0, and is referred to as {\em output}.
Each vertex of indegree 2 must be labeled by either $+$ or $\times$, thus
representing computation. Vertices are commonly referred to as {\em gates}
and edges as {\em arrows}.
\end{definition}

By interpreting the input gates either as constants or variables it is
easy to prove by induction that each arithmetic circuit naturally
represents a polynomial.

In this paper various subclasses of arithmetic circuits will be considered:
For {\em weakly skew} circuits we have the restriction that for every
multiplication gate, at least one of the incoming arrows is from a subcircuit
whose only connection to the rest of the circuit is through this incoming
arrow.
For {\em skew} circuits we have the restriction that for every
multiplication gate, at least one of the incoming arrows is from an
input gate.
For {\em formulas} all gates
(except output) have outdegree 1. Thus, reuse of partial
results is not allowed.

For a detailed description of various subclasses of arithmetic circuits,
along with examples, we refer to \cite{MP}.

\begin{definition}
The {\em size} of a circuit is the total number of {\em gates} in the circuit.
The {\em depth} of a circuit is the length of the longest path from an
input gate to the output gate.
\end{definition}

\subsection{Pathwidth and treewidth}\label{sec:pathwidthandtreewidth}

Since the definition of pathwidth is closely related to the definition
of treewidth (bounded pathwidth is a special case of bounded treewidth)
we also include the definition of treewidth in this paper.
Treewidth for undirected graphs is commonly defined as follows:

\begin{definition}
Let $G = \langle V,E \rangle$ be a graph. A $k$-tree-decomposition of $G$ is:
\begin{itemize}
\item[(i)] A tree $T = \langle V_T, E_T \rangle$.
\item[(ii)] For each $t \in V_T$ a subset $X_t \subseteq V$
of size at most $k + 1$.
\item[(iii)] For each edge $(u,v) \in E$ there is a $t \in V_T$ such that
$\lbrace u,v \rbrace \subseteq X_t$.
\item[(iv)] For each vertex $v \in V$ the set $\lbrace t \in V_T |
v \in X_t \rbrace$ forms a (connected) subtree of $T$.
\end{itemize}
The treewidth of $G$ is then the smallest $k$ such that there exists a
$k$-tree-decomposition for $G$.\\
A $k$-{\em path}-decomposition of $G$ is then a $k$-tree-decomposition where the
``tree'' $T$ is a path (each vertex $t \in V_T$ has at most one child in $T$).
\end{definition}

\begin{example}
Here we show that cycles
have pathwidth at most 2 by constructing a path-de\-com\-po\-si\-ti\-on of $G$
where each $X_t$ has size at most 3. Let $v_1, v_2, \ldots ,v_n$ be the
vertices of a graph $G$ which is a cycle.
The edges of $G$ are $(v_1,v_2),(v_2,v_3),
\ldots,(v_{n-1},v_n),(v_n,v_1)$. The vertex $v_1$ is contained in every
$X_t$ of the path-decomposition. Vertices $v_2$ and $v_3$ are contained
in $X_1$, vertices $v_3$ and $v_4$ are contained in $X_2$, and so on. Finally,
vertices $v_{n-1}$ and $v_n$ are contained in $X_{n-2}$.
This gives a path-decomposition of $G$ of width 2.
\end{example}

The pathwidth (treewidth) of a directed, weighted graph is naturally
defined as the
pathwidth (tree\-width) of the underlying, undirected, unweighted graph.
The pathwidth (treewidth) of an $(n \times n)$
matrix $M = (m_{i,j})$ is defined as
the pathwidth (treewidth) of the directed graph
$G_M = \langle V_M,E_M,w \rangle$ where
$V_M = \lbrace 1, \ldots , n \rbrace$, $(i,j) \in E_M$ iff
$m_{i,j} \neq 0$, and $w(i,j) = m_{i,j}$.
Notice that $G_M$ can have loops. Loops affect neither the pathwidth nor
the treewidth of $G_M$ but are important for the characterization of the
permanent polynomial.

\subsection{Cliquewidth, NLCwidth and m-cliquewidth}

Although there exists many algorithmic results for graphs of bounded
treewidth, there are still classes of ``trivial'' graphs that have
unbounded treewidth. Cliques are an example of such graphs.
Cliquewidth is a different notion of ``width'' for graphs, and it is more
general than treewidth since graphs of bounded treewidth have bounded
cliquewidth, but cliques have bounded cliquewidth and unbounded treewidth.

We recall the definitions of cliquewidth, NLCwidth and m-cliquewidth for
unweighted, undirected graphs.
Then we introduce the new notions of $W$-cliquewidth, $W$-NLCwidth and $W$-m-cliquewidth 
which are variants of the preceding ones for {\em weighted, directed} graphs.
These graph widths are all defined using terms over an universal algebra. When
we refer to parse-trees it means the parse-trees of these terms.

\begin{definition}[\cite{CER,CO}]
A graph $G$ has cliquewidth (denoted $cwd(G)$) at most $k$ iff there exists a set of source
labels ${\mathcal S}$ of cardinality $k$ such that $G$ can be constructed using
a finite number of the following operations (named clique operations):
\begin{itemize}
\item[(i)] $ver_a$, $a \in {\mathcal S}$
(basic construct: create a single vertex with label $a$).
\item[(ii)] $\rho_{a \rightarrow b} (H)$, $a,b \in {\mathcal S}$
(rename all vertices with label $a$ to have label $b$ instead).
\item[(iii)] $\eta_{a,b} (H)$, $a,b \in {\mathcal S}$, $a\neq b$
(add edges between all couples of vertices where one of them has label $a$
and the other has label $b$).
\item[(iv)] $H \; \oplus \; H'$ (disjoint union of graphs).
\end{itemize}
\end{definition}

\begin{example}
Using the clique algebra, the clique with four vertices $K_4$ is constructed by
the following term using only two source labels; $S = \{a,b\}$: 
$$\eta_{a,b} ((\rho_{a \rightarrow b} (\eta_{a,b} ((\rho_{a \rightarrow b} (\eta_{a,b}
(ver_a \; \oplus \; ver_b))) \; \oplus \; ver_a)))\; \oplus \; ver_a).$$
\end{example}

\begin{definition}[\cite{Wa}]
A graph $G$ has NLCwidth (denoted $w_{NLC}(G)$) at most $k$ iff there exists a set of source
labels ${\mathcal S}$ of cardinality $k$ such that $G$ can be constructed using
a finite number of the following operations (named NLC operations):
\begin{itemize}
\item[(i)] $ver_a$, $a \in {\mathcal S}$
(basic construct: create a single vertex with label $a$).
\item[(ii)] $\circ_R (H)$ for any mapping $R$ from ${\mathcal S}$ to
${\mathcal S}$ (for every source label $a \in {\mathcal S}$
rename all vertices with label $a$ to have label $R(a)$ instead).
\item[(iii)] $H \;\times_S \; H'$  for any $S\subseteq {\mathcal S}^2$
(disjoint union of graphs to which are added edges between
all couples of vertices $x \in H$ (with label $l_x$),
$y \in H'$ (with label $l_y$) having $(l_x,l_y) \in S$). 
\end{itemize}
\end{definition}

One important distinction
between cliquewidth and NLCwidth on one side and m-cliquewidth
(to be defined below) on the other side
is that in the first two each vertex is assigned exactly {\em one} label,
and in the
last one each vertex is assigned a \emph{set} of labels (possibly empty).

\begin{definition}[\cite{CT07}]
A graph $G$ has m-cliquewidth (denoted $mcwd(G)$) at most $k$ iff there exists a set of source
labels ${\mathcal S}$ of cardinality $k$ such that $G$ can be constructed using
a finite number of the following operations (named m-clique operations):
\begin{itemize}
\item[(i)] $ver_A$ (basic construct: create a single
vertex with a set of labels $A$, $A\subseteq {\mathcal S}$).
\item[(ii)] $H \;\otimes_{S,h,h'} \; H'$  for any $S\subseteq {\mathcal S}^2$
and any $h,h' : {\mathcal P}({\mathcal S}) \rightarrow
{\mathcal P}({\mathcal S})$
(disjoint union of graphs to which is added edges between
all couples of vertices $x \in H$, $y \in H'$ whose sets of labels $L_x,L_y$
contain a
couple of labels $l_x,l_y$ such that $(l_x,l_y) \in S$.
Then the labels of vertices from $H$ are changed via $h$ and the labels
of vertices from $H'$ are changed via $h'$). 
\end{itemize}
\end{definition}

It is stated in \cite{CT07} (a proof sketch of this result is given
in~\cite{CT07}, one of the inequalities is proven in \cite{Jo98}) that
$$mcwd(G)\leq wd_{NLC}(G) \leq cwd(G) \leq 2^{mcwd(G)+1}-1.$$
Hence, cliquewidth, NLC-width and m-cliquewidth
are equivalent with respect to boundedness.

\medskip

We have seen that the definition of pathwidth and treewidth for weighted
graphs straight forward was defined as the width of the underlying,
unweighted graph.
This is a major difference
compared to cliquewidth.
We can see
that if we consider non-edges as edges of weight 0, then every weighted graph
has a clique (which has bounded cliquewidth 2) as its underlying,
unweighted graph.

Our main motivation for studying bounded cliquewidth matrices is to obtain
efficient algorithms for evaluating polynomials like the permanent and
hamiltonian for such matrices.
For this reason, it is not reasonable
to define the cliquewidth of a weighted graph as the cliquewidth of the
underlying, unweighted graph, because then computing the permanent of a
matrix of cliquewidth 2 is as difficult as the general case.
Hence, we put restrictions on how weights are assigned to edges:
Edges added in the same operation between vertices having the same pair
of labels, will all have the same weight.

We now introduce the definitions of $W$-cliquewidth, $W$-NLCwidth and
$W$-m-cliquewidth.
We will consider simple, weighted, directed graphs where the weights are
in some set $W$. 
In the three following constructions, an
arc from a vertex $x$ to a vertex $y$
is only added by 
relevant operations if there is not already an arc from $x$ to $y$. 
The operations that differ from the unweighted case are
indicated by \textbf{bold} font.

\begin{definition}
A graph $G$ has $W$-cliquewidth (denoted $Wcwd(G)$) at most $k$ iff there exists a set of source
labels ${\mathcal S}$ of cardinality $k$ such that $G$ can be constructed using
a finite number of the following operations (named $W$-clique operations):
\begin{itemize}
\item[(i)] $ver_a$, $a \in {\mathcal S}$ 
(basic construct: create a single vertex with label $a$).
\item[(ii)] $\rho_{a \rightarrow b} (H)$, $a,b \in {\mathcal S}$
(rename all vertices with label $a$ to have label $b$ instead).
\item[\textbf{(iii)}] 
$\alpha_{a,b}^w (H)$, $a,b \in {\mathcal S}$, $a\neq b$, $w \in W$
(add missing arcs of weight $w$ from all vertices with label $a$
to all vertices with label $b$).
\item[(iv)] $H \; \oplus \; H'$ (disjoint union of graphs).
\end{itemize}
\end{definition}

\begin{definition}
A graph $G$ has $W$-NLCwidth (denoted $Wwd_{NLC}(G)$) at most $k$ iff there exists a set of source
labels ${\mathcal S}$ of cardinality $k$ such that $G$ can be constructed using
a finite number of the following operations (named $W$-NLC operations):
\begin{itemize}
\item[(i)] $ver_a$, $a \in {\mathcal S}$
(basic construct: create a single vertex with label $a$).
\item[(ii)] $\circ_R (H)$ for any mapping $R$ from ${\mathcal S}$ to
${\mathcal S}$
(for every source label $a \in {\mathcal S}$ rename all vertices with label
$a$ to have label $R(a)$ instead).
\item[\textbf{(iii)}] $H \;\times_S \; H'$  for any partial function
$S : {\mathcal S}^2 \times \{-1,1\} \rightarrow W $
(disjoint union of graphs to which are added arcs of weight $w$ 
for each couple of vertices $x\in H$, $y\in H'$ whose labels
$l_x,l_y$ are such that $S(l_x,l_y,s)=w$;
the arc is from $x$ to $y$ if $s=1$ and from $y$ to $x$ if $s=-1$). 
\end{itemize}
\end{definition}

\begin{definition}
A graph $G$ has $W$-m-cliquewidth (denoted $Wmcwd(G)$) at most $k$ iff there exists a set of source
labels ${\mathcal S}$ of cardinality $k$ such that $G$ can be constructed using
a finite number of the following operations (named $W$-m-clique operations):
\begin{itemize}
\item[(i)] $ver_A$ (basic construct: create a single
vertex with set of labels $A$, $A\subseteq {\mathcal S}$).
\item[\textbf{(ii)}] $H \;\otimes_{S,h,h'} \; H'$  for any partial function $S : {\mathcal S}^2 \times \{-1,1\} \rightarrow W $
and any
$h,h' : {\mathcal P}({\mathcal S}) \rightarrow {\mathcal P}({\mathcal S})$
(disjoint union of graphs to which is added missing arcs of weight $w$ 
for each couple of vertices $x\in H$, $y\in H'$
whose sets of labels $L_x,L_y$ contain
$l_x,l_y$ such that $S(l_x,l_y,s)=w$; 
the arc is from $x$ to $y$ if $s=1$ and from
$y$ to $x$ if $s=-1$. Then the labels of
vertices from $H$ are changed via $h$ and the labels of vertices
from $H'$ are changed via $h'$).
\end{itemize}
\end{definition}
In the last operation for $W$-m-cliquewidth,
there is a possibility that two (or more) arcs are added from a
vertex $x$ to a vertex $y$
during the same operation 
and then the obtained graph
is not simple. 
For this reason, we will consider
as well-formed terms only the terms (or parse-trees) where this does not occur.

The three preceding constructions of graphs can be extended to weighted
graphs with loops by adding the 
basic constructs $verloop_a^w$ or $verloop_A^w$ which creates a single
vertex with a loop of weight $w$ and label $a$ or set of labels $A$. 
If $G$ is a weighted graph (directed or not) with loops
and $Unloop(G)$ denotes the weighted graph (directed or not) obtained from $G$
by removing all loops, then one can easily show the following result.
\begin{itemize}
\item $Wcwd(G) = Wcwd(Unloop(G))$.
\item $Wwd_{NLC}(G) = Wwd_{NLC}(Unloop(G))$.
\item $Wmcwd(G) = Wmcwd(Unloop(G))$.
\end{itemize}

This justifies the fact that we overlook technical details for
loops in the proof of the following theorem. Theorem~\ref{weightedEquiv}
shows that the inequalities between the three widths are
still valid in the weighted case.
It justifies our definitions of cliquewidth for weighted graphs.
For the proof we collect
the ideas in~\cite{CT07,Jo98} and combine them with our definitions
for weighted graphs.

\begin{theorem}\label{weightedEquiv}
For any weighted graph $G$, $$Wmcwd(G)\leq Wwd_{NLC}(G) \leq Wcwd(G) \leq 2^{Wmcwd(G)+1}-1.$$
\end{theorem}
\proof
First inequality:

Let $G$ be a weighted graph of $W$-NLCwidth at most $k$ and $T$ be a parse-tree constructing $G$ 
with $W$-NLC operations on a set of source labels ${\mathcal S}$ of cardinality $k$.
We can consider without loss of generality that in $T$:
\begin{itemize}
\item[-] there are no two consecutive $\circ_R (H)$
operations, otherwise we can replace $T$ by $T'$ where the two consecutive nodes of $T$
with $\circ_R (H)$ and  $\circ_{R'} (H)$ operations on them have been replaced by one node $\circ_{R''} (H)$ ($R''=R'\circ R$).
\item[-] no $ver_a$ operation is followed by a $\circ_R (H)$ operation, otherwise we can replace $T$ by $T'$ where 
this two operations are replaced by $ver_b$ where $b=R(a)$.
\item[-] each $H \;\times_S \; H'$ operation is followed by exactly one $\circ_R (H)$ operation, otherwise 
we can add an $\circ_{Id} (H)$ operation if there is none ($Id$ is the identity function from ${\mathcal S}$
to ${\mathcal S}$).
\end{itemize} 
We can replace the $W$-NLC operation $ver_a$ by the $W$-m-clique operation $ver_{\{a\}}$,
and the consecutives $W$-NLC operation $H \;\times_S \; H'$ and $\circ_R (H)$ by the $W$-m-clique operation
$H \;\otimes_{S,h,h} \; H'$ where $h(\{a\})=\{R(a)\}, \forall a \in {\mathcal S}$.
It is clear that these replacements in $T$ will give a parse-tree constructing $G$ 
with $W$-m-clique operations on the same set of source labels ${\mathcal S}$ of cardinality $k$.
Hence, we have $Wmcwd(G)\leq Wwd_{NLC}(G)$.

Second inequality:

Let $G$ be a weighted graph of $W$-cliquewidth at most $k$ and $T$ be a parse-tree constructing $G$ 
with $W$-clique operations on a set of source labels ${\mathcal S}$ of cardinality $k$.
We can consider without loss of generality that in $T$:
\begin{itemize} 
\item[-] after a disjoint union operation $H \; \oplus \; H'$
all arcs in $G$ from $x \in H$ to $y\in H'$ (resp. from $y$ to $x$) are added 
between the disjoint union operation $H \; \oplus \; H'$ and the 
first following operation $O$ of disjoint union or renaming. Otherwise consider the first operation
$\alpha_{a,b}^w (H)$ after $O$ adding an arc between a vertex $x'$ from $H$ and a vertex $y'$ from $H'$. We can add an 
operation $\alpha_{a',b'}^w (H)$ before $O$ where $a'$(resp. $b'$) is the label in $H \; \oplus \; H'$ 
of the tail (resp. head) of the arc added by the operation $\alpha_{a,b}^w (H)$. 
\item[-] each operation $\alpha_{a,b}^w (H)$ add at least one arc.
\item[-] all $\alpha_{a,b}^w (H)$ operations are between a disjoint union operation $H \; \oplus \; H'$ and the 
first following operation $O$ of disjoint union or renaming.
\end{itemize}
We can replace the $W$-clique operation $ver_a$ by the $W$-NLC operation $ver_a$,
and the $W$-clique operation $\rho_{a \rightarrow b} (H)$ by the $W$-NLC operation $\circ_R (H)$
where $R(a)=b$ and $R(c)=c, \forall c \in {\mathcal S}, c\neq a$.
Finally each group consisting of a
$H \;\oplus \; H'$ $W$-clique operation and the following $\alpha_{a,b}^w (H)$ $W$-clique operations 
can be replaced by the $W$-NLC operation
$G \;\times_S \; G'$ where $S(a,b,1)=S(a,b,-1)=w$ if there is an $\alpha_{a,b}^w (H)$ operation
in the group.
It is clear that these replacements in $T$ will give a parse-tree constructing $G$ 
with $W$-NLC operations on the same set of source labels ${\mathcal S}$ of cardinality $k$.
Hence, we have $Wwd_{NLC}(G) \leq Wcwd(G)$.

Last inequality:

Let $G$ be a weighted graph of $W$-m-cliquewidth at most $k$ and $T$ be a parse-tree constructing $G$ 
with $W$-m-clique operations on a set of source labels ${\mathcal S}$ of cardinality $k$.
Let ${\mathcal S}'$ be a set of source labels of cardinality $2^{k+1}-1$, 
${\mathcal S}' ={\mathcal S}_l\sqcup {\mathcal S}_r \sqcup \{empty\} $ where $|{\mathcal S}_l|=|{\mathcal S}_r|=2^k -1$.
We define three bijections 
$l : {\mathcal P}({\mathcal S}) \backslash \emptyset \rightarrow {\mathcal S}_l$, 
$r : {\mathcal P}({\mathcal S}) \backslash \emptyset \rightarrow {\mathcal S}_r$,
and $u : {\mathcal S}_l\rightarrow {\mathcal S}_r$ such that $u(l(A))=r(A), \forall A \in {\mathcal P}({\mathcal S})$.
We will denote by $\rho_f$ a sequence of $\rho_{a\rightarrow b}$ $W$-clique operations
realizing a function $f$ from ${\mathcal S}'$ to ${\mathcal S}'$.
We associate to each function $S : {\mathcal S}^2 \times \{-1,1\} \rightarrow W $
a sequence $\alpha_S$ consisting of $\alpha_{l(A),r(B)}^w$ (resp. $\alpha_{r(B),l(A)}^w$) $W$-clique operations for all couples
$(a,b) \in {\mathcal S}^2, (A,B) \in ({\mathcal P}({\mathcal S}) \backslash \emptyset)^2$ such that $S(a,b,1)=w$ (resp. $S(a,b,-1)=w$), 
$a \in A$ and $b\in B$. 

We can replace the $W$-m-clique operation $ver_A$ by the $W$-clique operation $ver_{l(A)}$ if $A\neq \emptyset$ and
$ver_{empty}$ otherwise.
Each $W$-m-clique operation $H \otimes_{S,h,h'}  H'$ will be replaced by the following 
$W$-clique operations:
\begin{itemize} 
\item[-] apply $\rho_{u}$ to the subtree constructing $H'$.
\item[-] make a $H \;\oplus \; H'$ $W$-clique operation.
\item[-] apply $\alpha_S$.
\item[-] apply $\rho_{l\circ h \circ l^{-1}}$.
\item[-] apply $\rho_{l \circ h' \circ r^{-1}}$.
\end{itemize}

It is clear that these replacements in $T$ will give a parse-tree constructing $G$ 
with $W$-clique operations on the set of source labels ${\mathcal S}'$ of cardinality $2^{k+1}-1$.
Hence, we have $Wcwd(G) \leq 2^{Wmcwd(G)+1}-1$.
\qed

\subsection{Permanent and hamiltonian polynomials}

In this paper we take a graph theoretic approach to deal with permanent
and hamiltonian polynomials. The reason for this is that a natural way
to define pathwidth, treewidth or cliquewidth of a matrix $M$
is by the width of the graph $G_M$
(see Section \ref{sec:pathwidthandtreewidth}), also see e.g.~\cite{MM}.

\begin{definition}
A {\em cycle cover}
of a directed graph is a subset of the edges, such that
these edges form disjoint, directed cycles (loops are allowed).
Furthermore, each vertex in the
graph must be in one 
(and only one)
of these cycles. The {\em weight} of a cycle cover
is the product of weights of all participating edges.
\end{definition}

\begin{definition} \label{permdef}
The {\em permanent} of an $(n \times n)$ matrix $M = (m_{i,j})$ is the 
sum of weights of all cycle covers of $G_M$.
\end{definition}
The permanent of $M$ can also be defined by the formula
$$\per(M)=\sum_{\sigma \in S_n} \prod_{i=1}^n m_{i,\sigma(i)}.$$
The equivalence with Definition~\ref{permdef} is clear since any permutation 
can be written down as a product of disjoint cycles,
and this decomposition is unique.
The {\em hamiltonian} polynomial
$\ham(M)$ is defined similarly,
except that we only sum over
cycle covers
consisting of a {\em single} cycle (hence the name).

There is a natural way of representing polynomials by permanents.
Indeed, if the entries of $M$ are variables or constants from some field $K$, 
then $f=\per(M)$ is a polynomial with coefficients in $K$
(in Valiant's terminology,
$f$ is a projection of the permanent polynomial).
In the next sections we study the power of this representation in the case where
$M$ has bounded pathwidth or bounded cliquewidth.

\subsection{Connections between permanents and
sum of weights of perfect matchings}

Another combinatorial characterization of the permanent is by sum of
weights of perfect matchings in a bipartite graph. We will use this connection
to deduce results for the
permanent from results for the  sum of weights of perfect matchings and vice versa.

\begin{definition}
Let $G$ be a directed graph (weighted or not). We define the
{\em inside-outside graph} of $G$,
denoted $IO(G)$, as the bipartite, undirected graph (weighted or
not) obtained as follows: 
\begin{itemize}
\item split each vertex $u \in V(G)$ in two vertices $u^+$ and $u^-$;
\item  each arc $uv$ (of weight $w$) is replaced by an edge between $u^+$
and $v^-$ (of weight $w$). A loop on $u$ (of weight $w$) is replaced by an
edge between $u^+$ and $u^-$ (of weight $w$).
\end{itemize}
\end{definition}

It is well-known that the permanent of a matrix $M$ can be defined as
the sum of weights of all perfect matchings of $IO(G_M)$.
We can see that the adjacency matrix of $IO(G_M)$ is 
$\left(\begin{matrix}
0 & M \\
M^t & 0
\end{matrix} \right)
$.

\begin{lemma}\label{twIO}
If $G$ has treewidth (pathwidth) $k$, then $IO(G)$ has treewidth
(pathwidth) at most $2 \cdot k+1$.
\end{lemma}
\proof
Let $\langle T,(X_t)_{t\in V(T)} \rangle$ be a $k$-tree(path)-decomposition of $G$.
It is clear that $\langle T,(X'_t)_{t\in V(T)} \rangle$, where
$X'_t=\{u^+, u^- | u \in X_t\}$, is a tree(path)-decomposition
of $IO(G)$ of width $2 \cdot k+1$.
\qed

\begin{lemma}\label{cwIO}
If $G$ has $W$-cliquewidth $k$, then $IO(G)$ has $W$-cliquewidth at
most $2 \cdot k$.
\end{lemma}
\proof
Let $T$ be a parse-tree constructing $G$ with $W$-clique operations on a set of source labels ${\mathcal S}$ of cardinality $k$.
We can replace the $W$-clique operation $ver_a$ by the three operations $(ver_{a^+}) \;\oplus \; (ver_{a^-})$,
and the $W$-clique operation $\rho_{a \rightarrow b} (H)$ by the $W$-clique operations $\rho_{a^+ \rightarrow b^+} (H)$ and 
$\rho_{a^- \rightarrow b^-} (H)$.
Finally each $\alpha_{a,b}^w (H)$ $W$-clique operation 
can be replaced by the $\eta_{a^+,b^-}^w (H)$ $W$-clique operation.
It is clear that these replacements in $T$ will give a parse-tree constructing $IO(G)$ 
with $W$-clique operations on the set of source labels $\{a^+, a^- | a \in
{\mathcal S}\}$ of size $2 \cdot k$.
\qed


\section{Expressiveness of matrices of bounded pathwidth}

In this section we study the expressive power of permanents, hamiltonians
and perfect matchings of matrices of bounded pathwidth.
We will prove that in each case we capture exactly the families
of polynomials computed by polynomial size skew circuits of bounded width. 
A by-product of these proofs will be a proof of the equivalence between 
polynomial size skew circuits of bounded width and polynomial size \emph{weakly} 
skew circuits of bounded width. This equivalence can not be immediately
deduced from the already known equivalence between polynomial size skew 
circuits and polynomial size weakly skew circuits in the unbounded width 
case~\cite{To} (the proofs in~\cite{To} use a combinatorial characterization of
the complexity of the determinant as the sum of weights of $s,t$-paths in a
graph of polynomial size with distinguished vertices $s$ and $t$. The additional
difficulties to extend these proofs to circuits and graphs of bounded width
would be equivalent to the ones we deal with).
 We will then prove that skew circuits of bounded width
are equivalent to arithmetic formulas.

\begin{definition}
An arithmetic circuit $\varphi$ has {\em bounded width} $k \geq 1$
if there exists a finite
set of totally ordered layers such that:
\begin{itemize}
\item[-] Each gate of $\varphi$ is contained in exactly 1 layer.
\item[-] Each layer contains at most $k$ gates.
\item[-] For every non-input gate of $\varphi$ if that gate is in some layer
$n$, then both inputs to it are in layer $n+1$.
\end{itemize}
\end{definition}

\begin{theorem}\label{bwcircuitToPerm}
The polynomial computed by a weakly skew circuit of bounded width can be expressed as the
permanent of a matrix of bounded pathwidth.
The size of the matrix is polynomial in the size of the circuit.
All entries in the matrix are either 0, 1, constants of the polynomial,
or variables of the polynomial.
\end{theorem}
\proof
Let $\varphi$ be a weakly skew circuit of bounded width $k \geq 1$
and $l>1$ the number of layers in $\varphi$.
The directed graph $G$ we construct will have pathwidth at most $\left\lfloor
\frac{7\cdot k}{2}\right\rfloor -1$
(each bag in the path-de\-com\-po\-si\-ti\-on will contain at most
$\left\lfloor \frac{7\cdot k}{2}\right\rfloor$ vertices) and the number of bags in the
path-decomposition will be $l-1$.
$G$ will have two distinguished vertices $s$ and $t$, and the sum
of weights of all directed paths from $s$ to $t$ equals the value computed
by $\varphi$.
The vertex $s$ will be in all bags of the path-decomposition of $G$.

Since $\varphi$ is a weakly skew circuit we consider a decomposition of it
into disjoint subcircuits defined recursively as follows:
The output gate of $\varphi$ belongs to the {\em main subcircuit}. If a gate in the
main subcircuit is an addition gate,
then both of its input gates are in the
main subcircuit as well. If a gate $g$ in the main subcircuit is a
multiplication gate, then we know that at least one input to $g$ is
the output gate of a subcircuit which is disjoint from $\varphi$ except for its
connection to $g$. This subcircuit forms a {\em disjoint multiplication-input
subcircuit}. The other input to $g$ belongs to the main subcircuit.
If some disjoint multiplication-input subcircuit $\varphi'$
contains at least one multiplication gate, then we make a decomposition of
$\varphi'$ recursively.
Note that such a decomposition of a weakly skew circuit not necessarily is
unique (nor does it need to be), because {\em both} inputs to a
multiplication gate can be disjoint
from the rest of the circuit, and then any one of these two can be chosen as the
one that belongs to the main subcircuit.

Let $\varphi_0, \varphi_1, \dots, \varphi_d$ be the disjoint subcircuits obtained
in the decomposition ($\varphi_0$ is the main subcircuit). The graph $G$ will
have a vertex $v_g$ for every gate $g$ of $\varphi$ and $d+1$ additional
vertices $s=s_0, s_1, \dots, s_d$ ($t$ will correspond to $v_g$ where $g$ is the
output gate of $\varphi$). For every gate $g$ in the subcircuit $\varphi_i$, the
following construction will ensure that the sum of weights of directed paths
from $s_i$ to $v_g$ is equal to the value computed at $g$ in $\varphi$.

For the construction of $G$ we process the {\em decomposition} of $\varphi$
in a bottom-up manner.
Let sub\-cir\-cuit $\varphi_i$ be a leaf in the decomposition of $\varphi$
(so $\varphi_i$ consists solely of addition gates and input gates).
Assume that $\varphi_i$ is located in layers $top_i$ through $bot_i$
($1 \geq top_i \geq bot_i \geq l$) of $\varphi$.
First we add a vertex $s_i$ to $G$ in bag $bot_i - 1$, and
for each input gate with value $w$ in the bottom layer $bot_i$ of $\varphi_i$
we add a vertex to $G$ also in bag $bot_i - 1$ along with an edge of weight $w$
from $s_i$ to that vertex.
Let $n$ range from $bot_i - 1$ to $top_i$:
Add the already created vertex $s_i$ to bag $n-1$ and handle input gates
of $\varphi_i$ in layer $n$ as previously described.
For each addition gate of $\varphi_i$ in layer $n$
we add a new vertex to $G$ (which is added to bags $n$ and $n-1$ of
the path-decomposition of $G$). In bag $n$ we already have two vertices that
represent inputs to this addition gate, so we add edges of weight 1
from both of these to the newly added vertex.
The vertex representing the output gate of the circuit $\varphi_i$ is
denoted by $t_i$. The sum of weighted directed paths from $s_i$ to $t_i$
equals the value computed by the subcircuit $\varphi_i$.

Let $\varphi_i$ be a subcircuit in the decomposition of $\varphi$ that contains
multiplication gates. Addition gates and input gates
in $\varphi_i$ are handled as before.
Let $g$ be a multiplication gate in $\varphi_i$ in layer $n$ and
$\varphi_j$ the disjoint multiplication-input subcircuit
that is one of the inputs to $g$. We know that
vertices $s_j$ and $t_j$ already are in bag $n$, so we add an edge of
weight 1 from the vertex representing the other input to $g$ to
the vertex $s_j$, and an edge of weight 1 from $t_j$ to a newly
created vertex $v_g$ that represents gate $g$,
and then $v_g$ is added to bags $n$ and $n-1$.

For every $b$ ($1 \geq b \geq l-1$) we need to show that only a constant
number of vertices are added to bag $b$ during the entire process.
Every gate in layer $b$ of $\varphi$ is represented by a vertex, and these
vertices may all be added to bag $b$. Every gates in layer $b+1$ are also
represented by a vertex, and all of these are added to bag $b$
(because they are used as input here). So far we have at most $2 \cdot k$ gate
vertices in each bag.
In addition a number of $s_i$ vertices are also added to bag $b$.
For each subcircuit $\varphi_j$ that has a gate in layer $b$ or $b+1$,
we have the corresponding
$s_j$ vertex in bag $b$, so what remains is to show that at most
$\left\lfloor \frac{3 \cdot k}{2} \right\rfloor$ disjoint subcircuits
have a gate in layer $b$ or $b+1$. Each of these subcircuits are in exactly
one of the following 3 sets:
\begin{itemize}
\item[$C_1$:] Subcircuits that have a gate in layer $b$, but NONE
of them are multiplication gates.
\item[$C_2$:] Subcircuits that DO have a multiplication gate in layer $b$.
\item[$C_3$:] Subcircuits that have their root in layer $b+1$.
\end{itemize}
There are at most $\left\lfloor \frac{k}{2} \right\rfloor$ subcircuits
in the set $C_2$.
Otherwise, since two inputs to a multiplication gate are in different
subcircuits and since subcircuits in $C_2$ are disjoint
layer $b+1$ would contain at least $2 \cdot (\left\lfloor
\frac{k}{2} \right\rfloor + 1)$ gates and thus have width more than $k$.
By how subcircuits are constructed, all subcircuits in $C_3$ are considered
as the disjoint multiplication-input subcircuit of distinct
multiplication gates in layer $b$, so there are at least $| C_3 |$
multiplication gates in layer $b$. Since subcircuits in $C_1$ do NOT
have multiplication gates in layer $b$ we have that $|C_1| + |C_3| \leq k$.
Thus, at most $|C_1|+|C_2|+|C_3| \leq \left\lfloor \frac{3 \cdot k}{2}
\right\rfloor$ distinct subcircuits have their $s_i$ vertex added to bag $b$.

Note that in layer $1$ of $\varphi$ we just have the output gate. This gate
is represented by the vertex $t$ of $G$ which is in bag $1$ of the
path-decomposition. 

The sum of weights of all directed paths from $s$ to $t$
in $G$ can by induction
be shown to be equal to the value computed by $\varphi$.
The final step in the reduction to the permanent polynomial
is to add an edge of weight
$1$ from $t$ back to $s$ and loops of weight $1$ at all nodes different
from $s$ and $t$.
\qed

\medskip

The proof of Theorem~\ref{bwcircuitToPerm} can be modified to work for
the hamiltonian polynomial as well.
We adapt the idea used to show universality of the hamiltonian polynomial
in \cite{Mal}.
For the permanent polynomial
each bag in the path-decomposition contains at most $\left\lfloor \frac{7\cdot k}{2}\right\rfloor$
vertices; for each of these vertices we now need to introduce one extra
vertex in the same bag. In addition each bag must contain 2 more vertices in
order to establish a connection to adjacent bags in the path-decomposition.
In total each bag now contains at most $7 \cdot k + 2$ vertices.

\begin{theorem}\label{bwcircuitToMatch}
The polynomial computed by a weakly skew circuit of bounded width can be expressed as the
sum of weights of perfect matchings of a symmetric matrix of bounded pathwidth.
The size of the matrix is polynomial in the size of the circuit.
All entries in the matrix are either 0, 1, constants of the polynomial,
or variables of the polynomial.
\end{theorem}
\proof
It is a direct consequence of Theorem~\ref{bwcircuitToPerm} and Lemma~\ref{twIO}.
\qed

\medskip

Now we prove that the permanent, the hamiltonian, and the sum of weights of perfect
matchings of a bounded pathwidth graph can be expressed as a skew circuit of
bounded width.

\begin{theorem}\label{pathwidthHamToCircuit}
The hamiltonian of a matrix of bounded pathwidth can be expressed as a
skew circuit of bounded width.
The size of the circuit is polynomial in the size of the matrix.
\end{theorem}
\proof
Let $M$ be a matrix of bounded pathwidth $k$ and let $G_M$ be the underlying,
directed graph. Each bag in the path-decomposition of $G_M$ contains
at most $k+1$ vertices. We refer to one end of the path-decomposition as
the {\em leaf} of the path-decomposition and the other as the {\em root}
(recall that path-decompositions are special cases of tree-decompositions).

We process the path-decomposition of $G_M$ from the leaf towards the root.
The overall idea is the same as the proof of Theorem 5 in \cite{FKL}
-- namely to consider weighted partial path covers (i.e. partial covers consisting 
solely of paths) of 
subgraphs of $G_M$ that are induced by the path-decomposition of $G_M$.
During the processing of the path-decomposition of $G_M$ at every level distinct 
from the root, new partial path
covers are constructed by taking one previously generated partial path cover
and then add at most ${(k+1)}^2$ new edges, so all the multiplication gates
we have in our circuit are skew. For any bag in the path-decomposition
of $G_M$ we only need to consider a number of partial path covers that depends
solely on $k$, so the circuit we produce has bounded width.
At the root we add sets of edges to partial path covers to form
hamiltonian cycles.
\qed

\begin{theorem}\label{pathwidthMatchToCircuit}
The sum of weights of perfect matchings of a symmetric matrix of bounded pathwidth 
can be expressed as a skew circuit of bounded width.
The size of the circuit is polynomial in the size of the matrix.
\end{theorem}
\proof
Let $M$ be a symmetric matrix of bounded pathwidth $k$ and let $G_M$ be 
the underlying, undirected graph. Each bag in the path-decomposition of $G_M$ contains
at most $k+1$ vertices. 

We process the path-decomposition of $G_M$ from the leaf towards the root.
The proof is very similar to the proof of Theorem \ref{pathwidthHamToCircuit}
-- namely to consider weighted matchings of subgraphs of $G_M$ that are
induced by the matching of $G_M$.
During the processing of the matching of $G_M$ at every level 
distinct from the root, new matchings 
are constructed by taking one previously generated matching
and then add at most ${(k+1)}^2$ new edges, so all the multiplication gates
we have in our circuit are skew. For any bag in the path-decomposition
of $G_M$ we only need to consider a number of matchings that depends
solely on $k$, so the circuit we produce has bounded width.
At the root we sum only the weights of \emph{perfect} matchings to obtain the 
output of the circuit.
\qed

\begin{theorem}\label{pathwidthToCircuit}
The permanent of a matrix of bounded pathwidth can be expressed as a
skew circuit of bounded width.
The size of the circuit is polynomial in the size of the matrix.
\end{theorem}
\proof
It is a direct consequence of Theorem~\ref{pathwidthMatchToCircuit} and Lemma~\ref{twIO}.
\qed

\medskip

\begin{corollary}
A family of polynomials is computable by polynomial size skew circuits of bounded width
if and only if it is computable by polynomial size weakly skew circuits of bounded width.
\end{corollary}
\proof
It is trivial to see that a family of polynomials computed by polynomial size skew circuits of bounded width can 
be computed by polynomial size weakly skew circuits of bounded width.
Conversely, if a family of polynomials is computed by polynomial size weakly skew circuits of bounded width then
by Theorem \ref{bwcircuitToPerm} it can be expressed as the permanents of bounded pathwidth graphs
which can be computed by polynomial size skew circuits of bounded width according 
to Theorem~\ref{pathwidthToCircuit}.
\qed

\medskip

We need the following Theorem from \cite{BC} to prove the equivalence between polynomial size 
skew circuits of bounded width and polynomial size arithmetic formulas.

\begin{theorem}\label{formToLBS}
Any arithmetic formula can be computed by a 
linear bijection straight-line program of polynomial size that uses 
three registers.
\end{theorem}

Let $R_1,\dots ,R_m$ be a set of $m$ registers,
a linear bijection straight-line (LBS) program  is a vector 
of $m$ initial values given to the registers plus a 
sequence of instructions 
of the form
\begin{itemize}
\item[(i)] $R_j \leftarrow R_j + (R_i \times c)$, or
\item[(ii)] $R_j \leftarrow R_j - (R_i \times c)$, or
\item[(iii)] $R_j \leftarrow R_j + (R_i \times x_u)$, or
\item[(iv)] $R_j \leftarrow R_j - (R_i \times x_u)$,
\end{itemize}
where $1\leq i,j\leq m$, $i\neq j$, $1\leq u\leq n$, 
$c$ is a constant, and $x_1,\dots, x_n$ are variables ($n$ is the number of variables). 
We suppose without loss of generality that the value computed by the LBS program
is the value in the first register after all instructions have been executed.

\begin{theorem}\label{formulaEquiSBW}
A family of polynomials is computable by polynomial size skew circuits of
bounded width if and only if it is computable by a family of polynomial size arithmetic
formulas.
\end{theorem}
\proof
Let $(f_n)$ be a family of polynomials computable by polynomial size skew
circuits of bounded width, then
by Theorem \ref{bwcircuitToPerm} it can be expressed as the permanents of
bounded pathwidth graphs.
Since graphs of bounded pathwith have bounded treewidth, we know by Theorem 5
in~\cite{FKL} that 
it can be computed by a family of polynomial size arithmetic formulas.

Conversely, if $(f_n)$ is a family of polynomial size arithmetic formulas,
then by Theorem~\ref{formToLBS}, it is computable by linear bijection
straight-line programs
of polynomial size that use three registers. We will modify these programs 
to obtain equivalent skew circuits of width 6. At each step, the set of 
indices $\{i,j,k\}$ will be equal to $\{1,2,3\}$.

Suppose the initial values of the three registers are $r_1, r_2, r_3$, then 
the first layer of our skew circuit contains three input gates with the three
values $r_1, r_2, r_3$ along with two others inputs which will be defined according
to the next instruction in the straight-line program.

If the next instruction is $R_j \leftarrow R_j + (R_i \times U)$ where $U$ is 
a variable or a constant, then we assign the values $0$ and $U$ to the two input 
gates not already defined in the current layer $l$ and we create a new layer $l-1$ with 
three addition gates corresponding to $R_i,R_j,R_k$ whose inputs are the gate corresponding to $R_i$ 
(resp. $R_j,R_k$) in layer $l$ and the input with value $0$ in layer $l$. We also put 
a multiplication gate whose inputs are the gate corresponding to $R_i$ 
and the input with value $U$ in layer $l$. And we put again an input gate with value $0$.
Then we create a new layer $l-2$ with three addition gates corresponding to $R_i,R_j,R_k$ 
whose inputs are the gate corresponding to $R_i$ 
(resp. $R_j,R_k$) and the input with value $0$ for $i,k$
or the gate computing $(R_i \times U)$ for $j$ in layer $l-1$. We also put two others 
inputs which will be defined according
to the next instruction. 

If the next instruction is $R_j \leftarrow R_j - (R_i \times U)$, then we need 
to create one more layer than in the first case.  
We first assign the values $0$ and $U$ to the two input 
gates not already defined in the current layer $l$ and we create a new layer $l-1$ with 
three addition gates corresponding to $R_i,R_j,R_k$ whose inputs are the gate corresponding to $R_i$ 
(resp. $R_j,R_k$) in layer $l$ and the input with value $0$ in layer $l$. We also put 
a multiplication gate whose inputs are the gate corresponding to $R_i$ 
and the input with value $U$ in layer $l$. And we put again an input gate with value $0$ and another 
one with value $-1$.
Then we create an intermediate new layer $l-2$ with three addition gates corresponding to $R_i,R_j,R_k$ 
whose inputs are the gate corresponding to $R_i$ 
(resp. $R_j,R_k$) and the input with value $0$. We also put 
a multiplication gate whose inputs are the gate computing $(R_i \times U)$
and the input with value $-1$ in layer $l-1$. And we put again an input gate with value $0$.
Finally we create a new layer $l-3$ with three addition gates corresponding to $R_i,R_j,R_k$ 
whose inputs are the gate corresponding to $R_i$ 
(resp. $R_j,R_k$) and the input with value $0$ for $i,k$
or the gate computing $-(R_i \times U)$ for $j$ in layer $l-2$. We also put two others 
inputs which will be defined according
to the next instruction. 

In both cases, it is clear by induction that the three gates of the current layer
corresponding to $R_i,R_j,R_k$ are computing the values in these registers 
if we execute the instructions treated so far. Hence the result. 
\qed


\section{Expressiveness of matrices of bounded weighted cliquewidth}

In this section we study the expressive power of permanents, hamiltonians
and perfect matchings of matrices that have bounded weighted clique\-width.

We first prove that every arithmetic formula can be expressed as
the permanent, hamiltonian, or sum of weights of
perfect matchings of a matrix of bounded 
$W$-cliquewidth, using the results for the bounded pathwidth matrices and the
following lemma.

\begin{lemma}\label{pwcw}
Let $G$ be a weighted graph (directed or not) with weights in $W$.
If $G$ has pathwidth $k$, then $G$ has $W$-cliquewidth at most $k+2$.
\end{lemma}
\proof
Let $\langle T,(X_t)_{t\in V(T)} \rangle$ be a $k$-path-decomposition of $G$.
We refer to one end of the path-decomposition as
the {\em leaf} of the path-decomposition and the other as the {\em root}.
Let $G_t$ be the subgraph of $G$ induced by the vertices in bags below $X_t$.

We prove by induction on the height of $\langle T,(X_t)_{t\in V(T)} \rangle$
that every graph $G_t$ can be constructed by $W$-clique operations using at most
$k+2$ distinct labels. Moreover, at the end of this construction all vertices in
bag $X_t$ have distinct labels and all other vertices have a \emph{sink} label.

If $|V(T)|=1$ then $G$ has at most $k+1$ vertices. We can create them with $k+1$
distinct labels and add independently each edge between two vertices using
$W$-clique operations. 

Suppose $|V(T)|>1$, let $r$ be the root and $t$ be its child. By induction,
$G_t$ can be constructed  by $W$-clique operations using at most
$k+2$ distinct labels. For all vertex $v \in X_t \backslash X_r$, we add a renaming 
operation which gives \emph{sink} label to $v$ (this renaming operation renames
only $v$ since, by induction, $v$ has distinct label from other vertices). 
Since $|X_r| \leq k+1$ and all vertices in $V(G) \backslash X_r$ have \emph{sink}
label, we can create the vertices of $X_r \backslash X_t$ with distinct labels
and add them by disjoint union to the current construction. It is now clear that
all the vertices of $X_r$ have distinct labels thus we can add independently
each edge between two vertices. Hence the conclusion.
\qed

\begin{theorem}\label{formToPermClique}
Every arithmetic formula can be expressed
as the permanent of a matrix of $W$-cliquewidth at most $22$
and size polynomial in $n$,
where $n$ is the size of the formula. All entries in the matrix are
either 0, 1, constants of the formula, or variables of the formula.
\end{theorem}
\proof
Let $\varphi$ be a formula of size $n$. Due to the proof of
Theorem~\ref{formulaEquiSBW}, we know that it can be computed 
by a skew circuit of width 6 and size $O(n^{O(1)})$. 
Hence it is equal to the permanent of a graph of  size $O(n^{O(1)})$, pathwidth
at most $\left\lfloor \frac{7 \cdot 6}{2} \right\rfloor - 1 = 20$ by Theorem~\ref{bwcircuitToPerm},
and $W$-cliquewidth at
most $20+2=22$ by Lemma~\ref{pwcw}.
\qed

\medskip
For the hamiltonian the $W$-cliquewidth
becomes $((7 \cdot 6 +2 ) -1 ) +2 = 45$ instead.

\begin{theorem}\label{formToMatchClique}
Every arithmetic formula can be expressed
as the sum of weights of perfect matchings of a symmetric matrix of
$W$-cliquewidth at most $44$
and size polynomial in $n$,
where $n$ is the size of the formula. All entries in the matrix are
either 0, 1, constants of the formula, or variables of the formula.
\end{theorem}
\proof
It is a direct consequence of Theorem~\ref{formToPermClique} and Lemma~\ref{cwIO}.
\qed

\medskip

Alternatively we can modify the constructions of bounded treewidth
graphs expressing
formulas in~\cite{FKL}. These modifications require more work than the preceding
proofs but we obtain
smaller constants since we obtain graphs of $W$-cliquewidth at most 13/34/26
(instead of 22/45/44) whose permanent/hamiltonian/sum of weights of
perfect matchings are equal to formulas. The proofs of these constants are
given in the Appendix.

\medskip

Due to our restrictions on how weights are assigned in our
definition of bounded $W$-clique\-width it is not true that {\em weighted}
graphs
of bounded treewidth have bounded $W$-cliquewidth. In fact, if one tries to
follow the proofs in \cite{CO,CR} that show that graphs of bounded treewidth
have bounded cliquewidth, then one obtains that a weighted graph $G$ of treewidth $k$
has $W$-cliquewidth at most $3 \cdot (|W_G|+1)^{k-1}$ or $3 \cdot (\Delta + 1)^{k-1}$.
$W_G$ denotes the set of weights on the edges of $G$ and $\Delta$ is the maximum
degree of $G$.
Weighted trees still have bounded weighted cliquewidth (the bound is 3),
but we can show that there exists a family of weighted
graphs with treewidth 2 and unbounded $W$-cliquewidth~\cite{LT}.

\medskip

We now turn to the upper bound on the complexity of the permanent,
hamiltonian, and sum of weights of
perfect matchings of graphs of bounded weighted cliquewidth. We show that in all
three cases the complexity is at most the complexity of \vp.

The decision version of the hamiltonian cycle problem has been shown to
be polynomial time solvable in \cite{EGW} for matrices of bounded
cliquewidth. Here we extend these ideas in
order to compute the hamiltonian polynomial efficiently (in $\vp$)
for bounded $W$-m-cliquewidth matrices.

\begin{definition}
A {\em path cover} of a directed graph $G$ is a subset of
the edges of $G$, such that these edges form disjoint, directed,
non-cyclic paths in $G$.
We require that every vertex of $G$ is in (exactly) one path.
For technical reasons
we allow ``paths'' of length 0, by having paths that start and end in
the same vertex.
Such constructions do {\em not} have the same interpretation as a loop.
The {\em weight} of a path cover is the product of
weights of all participating edges (in the special case where there are no
participating edges the weight is defined to be 1).
\end{definition}

\begin{theorem}\label{hamilCliqueToCircuit}
The hamiltonian of an $n \times n$ matrix of bounded $W$-m-cliquewidth
can be expressed as a circuit of size $O(n^{O(1)})$
and thus is in $\vp$.
\end{theorem}
\proof
Let $M$ be an $n \times n$ matrix of bounded $W$-m-cliquewidth.
By $G$ we denote the underlying, directed, weighted graph for $M$.
The circuit is constructed based on the parse-tree $T$ for $G$.
By $T_t$ we denote the subtree of $T$ rooted at $t$ for some node $t \in T$.
By $G_t$ we denote the subgraph of $G$ constructed from the parse-tree $T_t$.

The overall idea is to produce a circuit that computes the sum of weights
of all hamiltonian cycles of $G$. To obtain this there will be
non-output gates that
compute weights of all path covers of all $G_t$ graphs,
and then we combine these subresults.
Of course, the total number of path covers can grow exponentially
with the size of $G_t$, so we will not ``describe'' path
covers directly by the edges participating in the covers.
Instead we describe a path cover of some $G_t$ graph by the labels
associated with the start- and end-vertices of the paths in the cover.
Such a description do not uniquely describe a path cover, because two different
path covers of the same graph can contain the same number of paths
and all these paths can have the same labels associated.
However, we do not need the weight of each individual path cover. If multiple
path covers of some graph $G_t$ share the same description, then we
just compute the sum of weights of these path covers.

For a leaf in the parse-tree $T$ of $G$ we construct a single gate of
constant weight 1, representing a path cover consisting of a single ``path''
of length 0, starting and ending in a vertex with the given labels.
Per definition this path cover has weight 1.

For an internal node $t \in T$ the grammar rule describes
which edges to add and how to relabel vertices. We obtain new path covers
by considering a path cover from the left child of $t$
and a path cover from the right child of $t$:
For each such pair of path covers consider all subsets of edges
added at node $t$, and for every subset of edges check if the addition
of these edges to the pair of path covers will result in a valid path cover.
If it does, then add a gate that computes the weight of this path cover,
by multiplying the weight of the left path cover, the weight of the
right path cover and the total weight of the newly added edges.
After all pairs of path covers have been processed, check if any of the
resulting path covers have the same description - namely that the number
of paths in some path covers are the same, and that these paths have the
same labels for start- and end-vertices.
If multiple path covers have the same description then add
addition gates to the circuit and produce a single gate which computes
the sum of weights of all these path covers.

For the root node $r$ of $T$ we combine path covers from the children
of $r$ to produce hamiltonian cycles, instead of path covers.
Finally, the output of the circuit is a summation of all gates computing
weights of hamiltonian cycles.

Proof of correctness:
The first step of the proof is by induction over the height of the parse-tree
$T$. We will show that for each non-root node $t$ of $T$
there is for every path cover description of $G_t$ a corresponding gate in the
circuit that computes the sum of weights of
all path covers of $G_t$ with that description.
For the base cases - leaves of $T$ - it is trivially true.

For the inductive step we consider two disjoint graphs that are being connected
with edges at a node $t$ of the parse-tree $T$.
Edges added at node $t$ are {\em only}
added in here, and not at any other nodes in $T$, so every path cover of
$G_t$ can be split into 3 parts: A path cover of $G_{t_l}$, a path cover of
$G_{t_r}$ and a polynomial number of edges added at node $t$.
Consider a path cover description along with all path covers of $G_t$
that have this description.
All of these path covers can be split into 3 such parts,
and by our induction hypothesis the weights of the path covers of $G_{t_l}$
and $G_{t_r}$ are computed in already constructed gates.

In order to complete the proof of correctness we have to handle the root $t$ of
$T$ in a special way. At the root we do not compute weights of
path covers, but instead compute weights of hamiltonian cycles. Every
hamiltonian cycle of $G$ can (similarly to path covers) be split into 3 parts:
A path cover of $G_{t_l}$, a path cover of $G_{t_r}$ and a polynomial number
of edges added at the root of $T$.
By our induction hypothesis all the needed weights are already computed.

The size of the circuit is polynomial since at each step the number of
path cover descriptions is polynomially bounded once the $W$-m-cliquewidth
is bounded.
\qed

\begin{theorem}\label{matchToVP}
The sum of weights of perfect matchings of an $n \times n$ symmetric
matrix of bounded $W$-NLCwidth
can be expressed as a circuit of size $O(n^{O(1)})$
and thus is in $\vp$.
\end{theorem}
\proof
Let $M$ be an $n \times n$ symmetric matrix of bounded $W$-NLCwidth.
By $G$ we denote the underlying, undirected, weighted graph for $M$.
The circuit is constructed based on the parse-tree $T$ for $G$.
By $T_t$ we denote the subtree of $T$ rooted at $t$ for some node $t \in T$.
By $G_t$ we denote the subgraph of $G$ constructed from the parse-tree
$T_t$. Let $k$ be the $W$-NLCwidth of $G$. We assume without loss of
generality that $T$ is a parse-tree on the set of labels $\{ a_1,\dots ,a_k\}$.

The overall idea is much similar to that of Theorem~\ref{hamilCliqueToCircuit},
namely to produce a circuit that computes the sum of weights
of all perfect matchings of $G$. To obtain this there will be
non-output gates that
compute weights of all matchings of all $G_t$ graphs,
and then we combine these subresults.
Of course, the total number of matchings can grow exponentially
with the size of $G_t$, so we will not ``describe'' matchings
directly by the edges participating in the covers.
Instead we describe a matching of some $G_t$ graph by the labels
associated to the uncovered vertices. More precisely, for each matching of $G_t$ and each
label $a$ we give the number of $a$-vertices which are not covered by the
matching. Such a description do not uniquely describe a matching, because two different
matchings of the same graph can have the same number of uncovered vertices which
have the same labels associated.
However, we do not need the weight of each individual matching. If multiple
matchings of some graph $G_t$ share the same description, then we
just compute the sum of weights of these matchings. It is clear that the number
of description needed is at most $n^k$.

For a leaf $ver_{a_i}$ in the parse-tree $T$ of $G$ we construct a
single terminal gate of
constant weight 1, representing an empty matching. The description associated to
this gate is $((a_1,0),\dots ,(a_i,1),\dots ,(a_k,0))$.

For an internal node $t \in T$ with operation $\circ_R (H)$ we just need to
change the description of terminal gates in the circuit contructed so far.
More precisely, if the description of the gate was
$((a_1,n_1),\dots ,(a_i,n_i),\dots ,(a_k,n_k))$ then it becomes
$$((a_1, \sum_{a_j \in R^{-1}(a_1)} n_j),\dots
,(a_i,\sum_{a_j \in R^{-1}(a_i)} n_j),\dots
,(a_k,\sum_{a_j \in R^{-1}(a_k)} n_j)).$$

For an internal node $t \in T$ with operation $H \;\times_S \; H'$
the grammar rule describes which edges to add. 
We first create a multiplication gate using the values of each couple of
terminal gates 
of the left child $l$ of $t$ and the right child $r$ of $t$.
It corresponds to the
weights of the disjoint unions of the matchings of $l$ and $r$. There is at most
$n^{2k}$ such gates. To each gate, we associate a left and right
description corresponding to the vertices from $l$ and $r$. Those gates are the
new terminal gates. 
We put the following total order
$a_1< a_2 < \dots <a_k$ on the labels and the corresponding
lexicographic order on the couples
$(a_i,a_j)$. We will consider that the edges added via $S$ are added by blocks
corresponding to a couple $(a_i,a_j)$ (All edges in the same block are added at
the same time) and that all blocks of edges are added sequentially in
lexicographic order. Thus we have at most $k^2$ steps of adding edges to
consider. Suppose $S(a_i,a_j)=w_{ij}$.
For the step corresponding to $(a_i,a_j)$ we obtain new matchings by
considering each terminal gate $g_0$. Let $((a_1,n_1),\dots
,(a_i,n_i),\dots ,(a_k,n_k))$ and $((a_1,n'_1),\dots
,(a_j,n'_j),\dots ,(a_k,n'_k))$ be the left and right description of $g_0$. 
Let $n_{min} = min\{n_i,n'j\}$. For all matching corresponding to $g_0$ 
and all $p$ between $0$ and $n_{min}$ we can obtain $\binom{n_i}{p} \cdot
\binom{n'_j}{p}$ matchings by adding $p$ edges of weight $w_{ij}$ between $p$
vertices among $n_i$ of $G_l$  and $p$ vertices among $n'_j$ of $G_r$. Hence,
for all $p\neq 0$ we add a multiplication gate with inputs $g_0$
and the constant $\binom{n_i}{p} \cdot \binom{n'_j}{p} \cdot (w_{ij})^p$.
This new gate $g_p$ has left and right
description $((a_1,n_1),\dots ,(a_i,n_i-p),\dots ,(a_k,n_k))$ and
$((a_1,n'_1),\dots ,(a_j,n'_j-p),\dots ,(a_k,n'_k))$. There are at most
$2\cdot n^{2k+1}$ such new gates since $p < n$. 
Finally we make an addition tree computing the addition of the gates $g_p$ which
have the same left and right description. Each such tree needs at most
$O((2k+2)\log(n))$ new gates and there are at most $2 \cdot n^{2k}$ trees.
The outputs of these
trees are the new terminal gates. When all the $k^2$ steps of adding edges are
done we compute the description of each terminal gate as the sum of
its left and right description then we put an addition tree computing the
addition of the terminal gates which have the same global description. 
The outputs of these trees are the new terminal gates.

Finally, we obtain the output of the circuit at the root node $r$ of $T$. It is
the output of the terminal gate with description
$((a_1,0),\dots ,(a_i,0),\dots ,(a_k,0))$.

Proof of correctness:
The first step of the proof is by induction over the height of the parse-tree
$T$. We will show that for each node $t$ of $T$
there is for every matching description of $G_t$ a corresponding gate in the
circuit that computes the sum of weights of
all matchings of $G_t$ with that description.
For the base cases - leaves of $T$ - it is trivially true.

For the inductive step we consider two disjoint graphs that are being connected
with edges at a node $t$ of the parse-tree $T$.
Edges added at node $t$ are {\em only}
added in here, and not at any other nodes in $T$, so every matching of
$G_t$ can be split into 3 parts: A matching of $G_{t_l}$, a matching of
$G_{t_r}$ and a polynomial number of edges added at node $t$.
Consider a matching description along with all matchings of $G_t$
that have this description.
All of these matchings can be split into 3 such parts,
and by our induction hypothesis the weights of the path covers of $G_{t_l}$
and $G_{t_r}$ are computed in already constructed gates.

The number of new gates added for each operation $H \;\times_S \; H'$
is at most $O(k^2 \cdot n^{2k+1})$.
Since the number of these operations is at most $n$,
we obtain a circuit of polynomial size.
\qed

\begin{theorem}
The permanent of an $n \times n$ matrix of bounded $W$-m-cliquewidth
can be expressed as a circuit of size $O(n^{O(1)})$
and thus is in $\vp$.
\end{theorem}
\proof
It is a direct consequence of Theorem~\ref{matchToVP} and Lemma~\ref{cwIO}.
\qed

\section{Acknowledgements}

Much of this work was done while U.~Flarup was visiting the ENS Lyon during
the spring semester of 2007.
This visit was partially made possible by funding from
Ambassade de France in Denmark,
Service de Coop\'eration et d'Action Culturelle,
Ref.:39/2007-CSU 8.2.1.


\section{Appendix}

\begin{theorem}\label{formToPermClique2}
Every arithmetic formula can be expressed
as the permanent of a matrix of $W$-cliquewidth at most $13$
and size polynomial in $n$,
where $n$ is the size of the formula. All entries in the matrix are
either 0, 1, constants of the formula, or variables of the formula.
\end{theorem}
\proof
Let $\varphi$ be a formula of size $n$.
Due to \cite{FKL} we know that $\varphi$ can be
expressed as the permanent of a matrix $M$ that has treewidth at most $2$ and
size at most $(n+1) \times (n+1)$.
Let $G$ be the underlying graph of $M$ and let $T = \langle V_T, E_T \rangle$
be the $2$-tree-decomposition of $G$. With only a linear increase in size
of $T$ we can assume that $T$ is a binary tree-decomposition.

Based on the tree-decomposition $T$ of $G$ we construct a graph $G'$
of bounded $W$-cliquewidth
such that (with slight abuse of notation) $per(G) = per(G')$.
A major difference between
grammars for bounded treewidth matrices and grammars for bounded clique\-width
matrices is that we cannot ``merge'' two vertices into a single vertex when
dealing with grammars for bounded cliquewidth matrices. As a consequence the
graphs $G$ and $G'$ will not be isomorphic, but there will be a 1 to 1
correspondence between their cycle covers.

For every non-loop edge $(u,v)$ of $G$ there can be multiple nodes $t \in V_T$
such that $u$ and $v$ both are in the set $X_t$. We say that an edge
$(u,v)$ of $G$ ``belong'' to a node $t \in V_T$, if $t$ is the node {\em
closest} to the root of $T$ where $u$ and $v$ both are in $X_t$ (for every edge
such a node is uniquely defined).

The general idea for the construction of $G'$ is as follows: We process $T$ in
a bottom-up manner. For a node $t \in V_T$ we first construct subgraphs
representing the children $l$ and $r$ of $t$, then we add the edges belonging
to $t$ using a special labeling scheme for the vertices. We do not have a label
in the grammar for each vertex of $G$ because this will not result in a
constant number of labels. Instead, since $|X_l| \leq 3$ and $|X_r| \leq 3$ we
use labels to represent vertices in $X_l$ and $X_r$ and reuse these labels
during the processing of $T$.

A vertex $v$ of $G$ is represented through multiple vertices in $G'$, but only
two of them are ``active'' at any time during the construction of $G'$: One
vertex of indegree 0 is managing edges leaving $v$ in $G$, and one vertex of
outdegree 0 is managing edges entering $v$ in $G$. Since $X_l$ and $X_r$ both
have size at most 3 we then need the following labels for this scheme: {\em
left-a-in, left-a-out, left-b-in, left-b-out, left-c-in} and {\em left-c-out}
(and 6 similar labels for {\em right}). In addition to that we also need a
{\em sink} label, giving a total of 13 labels needed to
construct $G'$. 

Processing $T$ to construct $G'$: For a leaf $t$ of $T$ we construct 6 vertices
(or 4, if $|X_t| = 2$), with the labels {\em left-a-in, left-a-out, left-b-in,
left-b-out, left-c-in} and {\em left-c-out} (assuming $t$ is the left child of
its parent). For non-loop edges belonging to node $t$, e.g. a directed edge
from the vertex represented with labels {\em left-b-in/out} to
the vertex represented with labels {\em left-a-in/out} of
weight $w$, we then add edges (actually just a single edge is added because
both of the labels are only assigned to one vertex of $G'$) from vertices
with label {\em left-b-out} to vertices with label {\em left-a-in} of weight
$w$. Next, if a vertex of $G$, e.g. the vertex represented by
{\em left-b-in/out}, is not present in $X_p$ ($p$ being the parent of $t$ in
$T$), then we add an edge of weight 1 from {\em left-b-in} to {\em left-b-out}.
Furthermore, if that vertex has a loop of weight $w$ we add an edge of weight
$w$ from {\em left-b-out} to {\em left-b-in}. In both cases we then rename
{\em left-b-out} and {\em left-b-in} to {\em sink}.

For an internal node $t \in V_T$ (including the root of $T$) we first consider
vertices of $G$ that are in both $X_l$ and $X_r$, e.g. {\em left-a-in/out} and
{\em right-b-in/out} represent the same vertex of $G$. 
We assume that $t$ is the left child of its parent in $T$.
We add a loop of weight 1 to each of {\em right-b-in} and {\em right-b-out}.
Then we add an edge of weight 1 from {\em right-b-in} to {\em left-a-in}
and an edge of weight 1 from {\em left-a-out} to {\em right-b-out}.
Then {\em right-b-in} and {\em right-b-out} are renamed to {\em sink}.
Next we add two vertices to $G'$ for every vertex in $X_t$ that are not in
$X_l$ nor $X_r$. There will be ``available'' {\em in/out} labels for these two
vertices, since in this case at least two other vertices were renamed to
{\em sink} during processing of each child of $t$. Next we
consider all edges of $G$ belonging to $t$. Assume there is a directed edge
from the vertex represented by {\em right-c-in/out} to the vertex represented
by {\em left-b-in/out} of weight $w$, then we
add an edge of weight $w$ from {\em right-c-out} to {\em left-b-in}.
Last, if a vertex of $G$, e.g. the vertex represented by {\em left-b-in/out},
is not present in $X_p$ ($p$ being the parent
of $t$ in $T$) or if $t$ is the root of $T$ then we add an edge of weight 1
from {\em left-b-in} to {\em left-b-out}. Furthermore, if that vertex has a
loop of weight $w$ we add an edge of weight $w$ from {\em left-b-out} to
{\em left-b-in}. In both cases we then rename {\em left-b-out} and
{\em left-b-in} to {\em sink}.

Proof of correctness: A vertex $v$ of $G$ is represented through two disjoint
sets of vertices in $G'$: One set of vertices managing edges entering $v$ in
$G$, and one set of vertices managing edges leaving $v$ in $G$. We denote these
sets of vertices in $G'$ as $v_{in}$ and $v_{out}$. A vertex of $G'$ belong to
$v_{in}$ if at some point during the processing of $T$ it were assigned an
{\em in} label which was representing $v$ in $G$. By our construction
it is clear that every vertex of $G'$ belong to either $v_{in}$ or $v_{out}$
for exactly 1 vertex $v$ of $G$, and the set $v_{in}$ form a
directed tree where all non-loop edges lead towards the root and have weight 1.
All non-root vertices in this tree have a loop of weight 1. The set $v_{out}$
has equivalent properties, with the exception that non-loop edges lead towards
the leaves instead of the root.

Now consider two vertices $u$ and $v$ of $G$ along with a directed edge of
weight $w$ from $u$ to $v$, and consider the trees $u_{out}$ and $v_{in}$ in
$G'$. At some point in the construction of $G'$ an edge of weight $w$ was added
from a vertex in $u_{out}$ to a vertex in $v_{in}$ in $G'$, so there is a path
of weight $w$ from the root of $u_{out}$ to the root of $v_{in}$ and all
vertices of
$u_{out}$ and $v_{in}$ not in this path have a loop of weight 1. So in a cycle
cover of $G$ where we include the edge from $u$ to $v$ we then have an
equivalent path in $G'$ and all remaining vertices in $u_{out}$ and $v_{in}$
are then covered by loops.
In order to ``continue'' the construction of the path
in $G'$ we then also have an edge of weight 1 from the root of $v_{in}$ to the
root of $v_{out}$. In order to simulate loops in cycle covers of $G'$ we have
added an edge from the root of $v_{out}$ back to the root of $v_{in}$ of same
weight as the loop in $G$. So a loop in $G$ corresponds to a cycle of length 2
in $G'$, and then all other nodes in both $v_{in}$ and $v_{out}$ are covered by
loops of weight 1.

It is then easy to verify that cycle covers in $G'$ are in bijection
with cycle covers of $G$ and the corresponding pairs of cycle covers
have same weight. Finally,
note that between any two vertices of $G'$ there is at most 1 edge so we can
find a matrix $M'$ such that the underlying graph of $M'$ is equivalent to
$G'$ and then $per(M') = per(M)$.
\qed

\begin{theorem}\label{formToHamClique2}
Every arithmetic formula can be expressed as the hamiltonian of
a matrix of $W$-cliquewidth at most 34 and size polynomial in $n$, where $n$ is
the size of the formula. All entries in the matrix are either 0, 1, or
constants of the formula, or variables of the formula.
\end{theorem}
\proof
Let $\varphi$ be a formula of size $n$.
Due to \cite{FKL} we know that $\varphi$ can be expressed as the
hamiltonian of a matrix $M$ that has treewidth at most $6$ and size at most
$(2n+1) \times (2n+1)$. Let $G$ be the underlying, weighted, directed graph
for the matrix $M$ and let $T = \langle V_T, E_T \rangle$ be the binary
$6$-tree-decomposition of $G$.
With only a linear increase in size of $T$ we can assume that $T$ is a
binary tree-decomposition.

The overall idea is the same as in Theorem~\ref{formToPermClique2}
- namely to process the tree-decomposition $T$ of $G$.
Since all $|X_t| \leq 7$ in this tree-decomposition we instead need
at least $2 \cdot 14+1=29$ labels during the processing
of $T$ to construct $G'$.

However, if we just use the exact same idea as in
Theorem~\ref{formToPermClique}, then for every cycle cover in the produced
graph many vertices are covered through loops.
Instead of introducing such loops we ``eliminate'' them using the
same idea as in \cite{Mal} used for showing universality of the hamiltonian
polynomial.

We need 5 additional labels for this construction:
{\em left-h1, left-h2, right-h1,
right-h2} and {\em temp}, for a total of 34 labels.
For a leaf $t$ of $T$ we start
the processing of $t$ by constructing two vertices and label them
{\em left-h1} and {\em left-h2} (assuming $t$ is the left child of
its parent in $T$),
and add an edge of weight 1 from {\em left-h1} to {\em left-h2}.
Remaining processing of $t$ is done as before.

For an internal node $t$ of $T$ we first add an edge of weight 1 from
{\em left-h2} to {\em right-h1}, rename {\em left-h2} and {\em right-h1}
to {\em sink}, and rename {\em right-h2} to {\em left-h2} (assuming $t$ is the
left child of its parent in $T$).
Some vertices, e.g. the vertex with label {\em right-c-in},
may have a loop added during the processing of $t$.
Instead of adding such a loop we do the following:
Add a new vertex with label {\em temp},
add an edge of weight 1 from {\em left-h2} to {\em right-c-in},
add an edge of weight 1 from {\em right-c-in} to {\em temp},
add an edge of weight 1 from {\em left-h2} to {\em temp},
rename {\em left-h2} to {\em sink},
rename {\em temp} to {\em left-h2}.
Remaining processing of $t$ is done as before.

When we reach the root $r$ of $T$ we consider any vertex of $X_r$, e.g. the
vertex represented by labels {\em left-a-in/out}. In the final step, instead of
adding an edge of weight 1 from {\em left-a-in} to {\em left-a-out}, we add an
edge of weight 1 from {\em left-a-in} to {\em left-h1}
and an edge of weight 1 from {\em left-h2} to {\em left-a-out}.
Now, for every hamiltonian cycle of $G$ we break
up the equivalent cycle of $G'$ and visit any remaining vertices of $G'$ along
a path of total weight 1.
\qed

\begin{theorem}\label{formToMatchClique2}
Every arithmetic formula can be expressed
as the sum of weights of perfect matchings of a symmetric matrix of $W$-cliquewidth at most $26$
and size polynomial in $n$,
where $n$ is the size of the formula. All entries in the matrix are
either 0, 1, constants of the formula, or variables of the formula.
\end{theorem}
\proof
It is a direct consequence of Theorem~\ref{formToPermClique2} and
Lemma~\ref{cwIO}.
\qed

\end{document}